\pdfoutput=1 

\documentclass[twocolumn,tighten,times]{aastex631}

\usepackage{color}
\usepackage{url}
\usepackage{hyperref}
\usepackage{xspace}
\usepackage{soul}

\shorttitle{X-ray polarization of PSR~B1259--63}
\shortauthors{Kaaret et al.}

\begin{document}

\title{Magnetic field geometry of the gamma-ray binary PSR~B1259--63 revealed via X-ray polarization}

\author[0000-0002-3638-0637]{Philip Kaaret}
\affiliation{NASA Marshall Space Flight Center, Huntsville, AL 35812, USA}

\author[0000-0002-7150-9061]{Oliver J. Roberts}
\affiliation{Science and Technology Institute, Universities Space Research Association, Huntsville, AL 35805, USA}

\author[0000-0003-4420-2838]{Steven R. Ehlert}
\affiliation{NASA Marshall Space Flight Center, Huntsville, AL 35812, USA}

\author[0000-0002-2954-4461]{Douglas A. Swartz}
\affiliation{Science and Technology Institute, Universities Space Research Association, Huntsville, AL 35805, USA}

\author[0000-0002-5270-4240]{Martin C. Weisskopf}
\affiliation{NASA Marshall Space Flight Center, Huntsville, AL 35812, USA}

\author[0000-0001-9200-4006]{Ioannis Liodakis}
\affiliation{NASA Marshall Space Flight Center, Huntsville, AL 35812, USA}
\affiliation{Institute of Astrophysics, Foundation for Research and Technology-Hellas, GR-70013 Heraklion, Greece}

\author[0000-0001-7163-7015]{M. Lynne Saade}
\affiliation{Science and Technology Institute, Universities Space Research Association, Huntsville, AL 35805, USA}

\author[0000-0002-1868-8056]{Stephen L. O'Dell}
\affiliation{NASA Marshall Space Flight Center, Huntsville, AL 35812, USA}

\author[0000-0002-4945-5079]{Chien-Ting Chen}
\affiliation{Science and Technology Institute, Universities Space Research Association, Huntsville, AL 35805, USA}


\begin{abstract}

Some X-ray binaries containing an energetic pulsar in orbit around a normal star accelerate particles to high energies in the shock cone formed where the pulsar and stellar winds collide. The magnetic field geometry in the acceleration region in such binaries is unknown. We performed the first measurement of the polarization of the X-ray synchrotron emission from a gamma-ray emitting binary system. We observed PSR~B1259--63 with the Imaging X-ray Polarimetry Explorer (IXPE) during an X-ray bright phase following the periastron passage in June 2024. X-ray polarization is detected with a polarization degree of $8.3\% \pm 1.5\%$ at a significance of $5.3\sigma$. The X-ray polarization angle is aligned with the axis of the shock cone at the time of the observation. This indicates that the predominant component of the magnetic field in the acceleration region is oriented perpendicular to the shock cone axis.

\end{abstract}


\section{Introduction}
\label{sec:intro}

Understanding particle acceleration in relativistic shocks is a key goal in high energy astrophysics. Pulsars lose rotational energy primarily through the generation of relativistic winds of electromagnetic energy and electron-positron pairs \citep{Curtis1982}. Pulsars in isolation produce pulsar wind nebulae (PWNe) that emit non-thermal emission from radio up to the TeV band produced by relativistic electrons accelerated at the pulsar wind termination shock \citep{Gaensler2006}. 

In contrast, the winds from pulsars in stellar binaries interact with the stellar outflows and particle acceleration occurs in a time-varying standing shock \citep{Tavani1997}. There are three stellar binary systems that are known to contain energetic pulsars and emit TeV gamma-rays indicating the presence of highly energetic particles \citep{HESSB1259,VERJ2032,Weng2022}. In these systems, the companion is an early-type star (spectral type O or B) that produces a stellar wind and an outflowing circumstellar disk. Unpulsed, non-thermal emission is observed from the radio band up to TeV energies.

The X-ray emission in both PWNe and TeV binaries is due to synchrotron radiation by the energetic particles. Therefore, X-ray polarization can be used to probe the configuration of the magnetic field near the acceleration region. We observed the TeV binary PSR~B1259--63 with the Imaging X-Ray Polarimetry Explorer \citep[IXPE,][]{Weisskopf2022} during an X-ray bright phase shortly after its periastron passage on 30 June 2024 (MJD 60491.57). We describe the binary system and the IXPE observations in Section~\ref{sec:obs}. Our analysis and the results are presented in Section~\ref{sec:results}. We conclude with a discussion of the implications for our understanding of the magnetic field geometry in TeV binaries in Section~\ref{sec:discuss}.

\section{IXPE Observations of PSR~B1259--63}
\label{sec:obs}

PSR~B1259--63, hereafter B1259, is a neutron star with a spin period of $P = 47.76$~ms and a spin-down power of $\approx 8 \times 10^{35}$~erg/s \citep{Johnston1994} in a binary system at a distance of 2.6~kpc \citep{MillerJones2018}. The companion star is a massive, emission line star with spectral type O9.5Ve, but often referred to as a Be star, that produces a roughly isotropic wind and a dense equatorial outflow which forms a stellar disk that is inclined to the orbital plane \citep{Dubus2013}. The orbital period is 1236.724526~days \citep{Shannon2014}. The pulsar plunges through the companion star's circumstellar disk twice each orbit, about 16 days before and 13 days after periastron. The timing is estimated from the disappearance of the radio pulsations around periastron, thought to be caused by free-free absorption in the stellar disk \citep{Johnston1999}. The X-ray emission strongly brightens as the pulsar passes through the disk \citep{Chernyakova2009}. The X-ray spectrum is well described by a powerlaw as expected for synchrotron emission \citep{Grove1995}. Very high energy (VHE) gamma-ray emission is also seen after disk crossings \citep{HESSB1259}.

The IXPE observations occurred shortly after the disk passage following periastron in 2024, from 14 July 2024 (MJD 60505.417; Bastille Day) to 29 July 2024 (MJD 60520.163). IXPE has three X-ray telescopes, each of which consists of a Mirror Module Assembly \citep[][MMA]{Ramsey2022} and a detector unit (DU) containing a polarization-sensitive gas-pixel detector (GPD) that measures linear X-ray polarization by imaging the tracks of photoelectrons produced by incident X-rays \citep{Costa2001,Soffitta2021,Baldini2021}. The IXPE photon event list was processed to reduce the instrumental background using the rejection algorithm described in \citet{DiMarco2023}. We also removed times of high background by filtering on the DU counting rates in the 2-8~keV band, excising intervals when the count rate in any DU in a 240~s time bin exceeded the mean rate in that DU by more than 3 times the standard error on the rate measurement. After filtering, the total livetime was 794.9, 794.5, and 794.6~ks for DU1, DU2, and DU3, respectively. We extracted source counts from a circular region with a radius of $60\arcsec$ which was found to optimize the signal to background ratio and background counts from a concentric annulus with radii of $150\arcsec$ and $300\arcsec$ \citep{DiMarco2023}. The source spectrum is above the background spectrum, scaled for the region sizes, in the 2-6~keV band; we use that energy band for the subsequent IXPE analysis. We used tools in both HEASoft version 6.33.2 and \textsc{ixpeobssim} version 31.0.1 \citep{Baldini2022} for analysis. We used version 20240125 of the GPD response files and version 20231201 of the MMA response files. All uncertainties are reported at 68.27\% confidence.


\begin{figure}[tb]
\centerline{\includegraphics[width=3.25in]{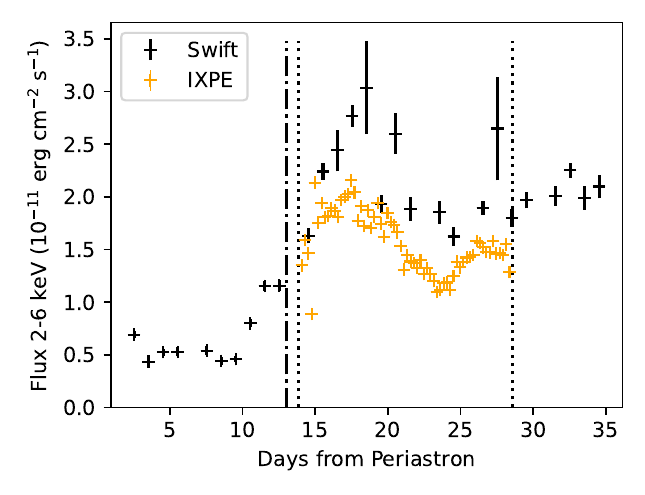}}
\caption{IXPE and Swift X-ray light curves in the 2-6~keV band. The vertical dash-dotted line shows when the pulsar crosses the stellar disk. The vertical dotted lines show the start and end of the IXPE observation.}
\label{fig:swift_lc}
\end{figure}

\begin{figure} 
\centerline{\includegraphics[width=2.2in]{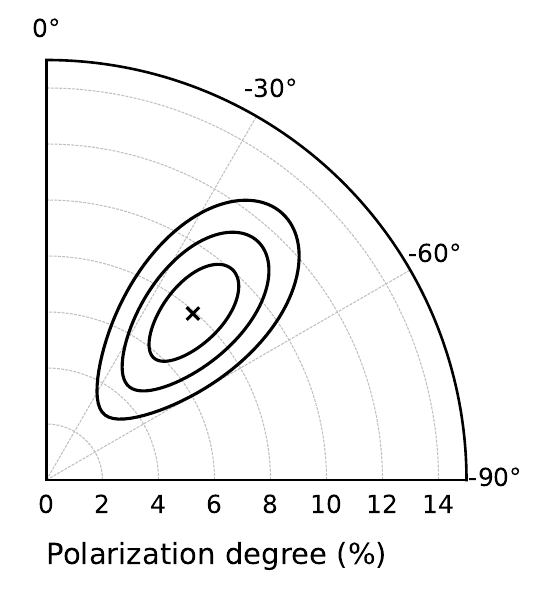}}
\caption{Contour plots of the polarization degree and angle without correction for background. The $\times$ marks the measured values. The contours show the 68.27\%, 95.45\%, and 99.73\% confidence intervals for $\chi^2$ on 2 degrees of freedom.} 
\label{fig:pcube_polar}
\end{figure}

\section{Results}
\label{sec:results}

Figure~\ref{fig:swift_lc} shows the X-ray light curve measured with IXPE and with the Swift X-Ray Telescope \citep[XRT;][]{Burrows2005}. The time bins are 1~day for the XRT and 19.6~ks for IXPE. Fluxes in the 2-6~keV band were estimated from IXPE count rates in the 2-6~keV band and XRT count rates in the 0.3-10~keV band assuming an absorbed powerlaw spectrum with an absorption column density of $7 \times 10^{21} \rm \, cm^{-2}$ \citep{Chernyakova2023} and a photon index of 1.6 as measured in the IXPE spectropolarimetric analysis below. We used the Swift-XRT products webpage at the University of Leicester.

Figure~\ref{fig:pcube_polar} shows the polarization of the events from the source region for the full observation as measured using the model-independent \texttt{pcube} algorithm with no additional correction for background. The polarization signal of each event was weighted based on the capability of the electron track characterization algorithm to correctly reconstruct the original photoelectron direction. The observed polarization degree (PD) is $7.9\% \pm 1.4\%$. The probability that the signal is a random fluctuation from an unpolarized source is $1.0 \times 10^{-7}$. This is equivalent to a $5.3 \sigma$ confidence level and indicates a secure detection of X-ray polarization.  The electric vector position angle (EVPA) is $-41\fdg3 \pm  5\fdg0$ and is measured anticlockwise from north in the equatorial coordinate system.

Correcting for background utilizing the additive nature of the Stokes parameters, we estimate the PD of the source as $8.31\% \pm 1.45\%$ and the EVPA as $-41\fdg7 \pm  5\fdg0$. As expected for unpolarized backgrounds, the EVPA is consistent with the results without background subtraction. 


We also performed a spectropolarimetric analysis using Xspec \citep{xspec}. We modeled the spectrum using an absorbed powerlaw with constant polarization multiplied by a constant allowed to vary between the three DUs. The column density for the \texttt{tbabs} absorption model using the \citet{Wilms2000} abundances was fixed to $7 \times 10^{21}\,\rm cm^{-2}$ \citep{Chernyakova2023}. The normalization for DU1 was fixed to unity and those for the other DUs were allowed to vary. We used a weighted polarization analysis. We find a good fit ($\chi^2/\rm dof = 829.6/885$) with a photon index of $\Gamma = 1.602 \pm 0.013$. The observed flux is $1.6 \times 10^{-11}\,\rm \, erg\,cm^{-2}\,s^{-1}$ in the 2--6~keV band. The EVPA is $-43\fdg8 \pm  4\fdg7$, which is consistent with the \texttt{pcube} results. The PD is $8.32\% \pm 1.37\%$, which is consistent with the background-subtracted \texttt{pcube} result.

We searched for variability in the polarization signal by dividing the observation in half and using \texttt{pcube} to estimate the polarization. In the first half, polarization is detected at a confidence level of $1 - 1.8 \times 10^{-8}$ with a PD of $11.11\% \pm 1.86\%$ and an EVPA of $-41\fdg1 \pm 4\fdg8$. In the second half, the confidence level drops to 0.81 with a PD of $3.83\% \pm 2.11\%$ and an EVPA of $-42\fdg1 \pm 15\fdg8$. To test the significance of the change in PD assuming constant PA, we rotate the Stokes parameters for each observation by the PA measured for the full observation and compare the Stokes Q values. We find that $\Delta Q = 0.0728 \pm 0.0281$ corresponding to a confidence level of 99\%. Thus, there is probable, but inconclusive, evidence for variability.



\begin{figure}[tb]
\centerline{\includegraphics[width=3.7in]{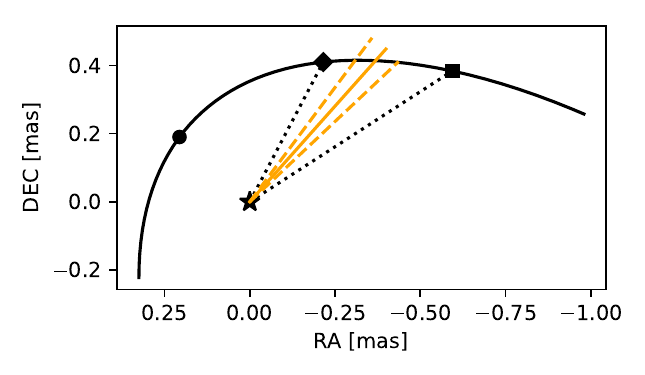}}
\caption{Orbital motion of PSR~B1259-63. The pulsar moves clockwise around the orbit shown as the solid curve. Marked along the orbit, in time order, are periastron (circle), start of IXPE observation (diamond), observation end (square). The star shows the system barycenter. The X-ray EVPA is shown as the solid orange line radiating from the barycenter with its uncertainty shown as the two dashed orange lines. The two black dotted lines show the position angle of the pulsar and, thus, the shock cone at the start and end of the IXPE observation.}
\label{fig:orbit}
\end{figure}

\section{Discussion}
\label{sec:discuss}

The IXPE measurements of the X-ray polarization of PSR~B1259-63 provide new information about the magnetic field in the shock region where the relativistic wind from the pulsar interacts with the outflows from Be star. The pulsar wind termination shock is roughly symmetric around the axis between the pulsar and the star. The geometry depends on the ratio, $\eta$, of the momentum flux density of the pulsar wind to that of the Be star outflow at the shock location. The IXPE observations occur just after a disk passage when $\eta \approx 5 \times 10^{-2}$  \citep{Bogovalov2008}. With this momentum ratio, the shock has a cone-like shape opening towards the pulsar with a bow shock at the apex curved away from the Be star.

Figure~\ref{fig:orbit} shows a portion of the orbit of B1259 around the times of the IXPE observation. The pulsar orbit was measured via highly accurate radio timing and astrometric observations \citep{Shannon2014,MillerJones2018}. The position angle of the pulsar relative to the center of mass, which sets the shock-cone position angle, varies from $-27\fdg7$ at the observation start to $-57\fdg1$ at the stop. The EVPA of $-41\fdg7 \pm  5\fdg0$ lies between the shock cone position angles at the start and stop of the observation and is consistent with the average shock cone position angle during the observation. We re-analyzed the data by rotating the Stokes parameters for each event according to the pulsar/shock-cone position angle at the time of the event. However, this did not increase the PD nor the detection significance.

For synchrotron emission, the EVPA is perpendicular to the magnetic field. Thus, the dominant component of the magnetic field is perpendicular to the shock axis of symmetry. A random magnetic field geometry would produce zero net polarization which is inconsistent with the PD observed. A poloidal field geometry, with a small shock-cone opening angle, would produce a net magnetic field almost parallel to the shock-cone axis \citep{Xingxing2021} and, thus, an EVPA perpendicular to that observed. The IXPE results are consistent with a toroidal magnetic field at the termination shock. However, the observations provide insufficient information to uniquely identify the field geometry as toroidal.

\citet{Dubus2015} performed relativistic hydrodynamical simulations of shocks in TeV binaries. The simulations were performed for LS~5039. However, the shock geometry and flow dynamics, after scaling for the orbital separation ($d$), should be reasonably well matched to B1259 since the selected momentum flux ratio was taken as $\eta = 0.1$ similar to that expected during the IXPE observations of B1259. In the simulations, most of the X-ray emission arises from a compact region at the apex with an extent of $\sim$ 0.5$d$. The orbital separation during the IXPE observation was $\sim$1000~light-seconds, thus the emission/acceleration region would extend $\sim$500~light-seconds. The small extent is due to the short timescales for acceleration and radiative losses. These timescales are expected to be $\sim$200~s for B1259 near periastron \citep{Tavani1997}. X-ray variability has been detected from B1259 on timescales of 3~ks around when the pulsar passes through the stellar disk \citep{Chernyakova2009,Uchiyama2009}. Very rapid X-ray variability has been seen from the TeV binary LS~I~+61$\arcdeg$~303, which also contains a pulsar but has a smaller orbital separation, on timescales down to a few seconds \citep{Smith2009}.

The termination shock is predominantly orthogonal to the symmetry/flow axis in the compact region where the simulations suggest that particle acceleration and X-ray emission occurs. Thus, the magnetic field in the acceleration region is orthogonal to the flow, i.e. the acceleration occurs in a perpendicular shock. \citet{Sironi2013} studied particle acceleration in relativistic shocks as a function of the magnetization, $\sigma$, which is the ratio of magnetic to kinetic energy in the flow. They found that for perpendicular shocks efficient acceleration occurs only for low magnetization, $\sigma < 10^{-3}$. This may suggest that the magnetization in the pulsar wind is already low at the termination shock in PSR~B1259-63 which lies at $\sim 2 \times 10^4 R_L$, where the light cylinder radius $R_L = cP/2 \pi$.


The PD for synchrotron emission depends on the particle spectral index and the level of uniformity of the magnetic field. The maximum polarization for synchrotron radiation given the measured X-ray photon index is 71\%. The PD of $8.3\% \pm 1.4\%$ observed in B1259 is significantly lower. The ratio of the PD to the maximum polarization is roughly given by the ratio of the energy in the uniform magnetic field component to that in the total field  \citep{Burn1966}; this would suggest that $\approx 12$\% of the field is in the uniform component. However, the observed PD is reduced relative to the intrinsic value if the shock-cone axis is not perpendicular to the line of sight \citep{Laing1980,Xingxing2021}. The inclination of the shock-cone axis varies from $74\arcdeg$ to $66\arcdeg$ over the observation \citep{MillerJones2018}. Also, the curvature of the bow shock could reduce the observed PD relative to the local PD at each position along the shock. Thus, the fraction of the field in the uniform component should be considered as a lower bound.

In conclusion, IXPE observations of PSR~B1259-63 have revealed, for the first time, the magnetic field geometry in a gamma-ray binary. The magnetic field in the acceleration region is predominantly aligned perpendicular to the shock-cone axis. This new information on the magnetic field geometry should inform future studies of relativistic shock acceleration.

\begin{acknowledgments}

We thank the anonymous referee for comments that improved the paper. We thank Allyn Tennant and Brian Ramsey for useful discussions. 
The Imaging X-ray Polarimetry Explorer (IXPE) is a joint US and Italian mission.  The US contribution is supported by the National Aeronautics and Space Administration (NASA) and led and managed by its Marshall Space Flight Center (MSFC), with industry partner Ball Aerospace (now, BAE Systems).  The Italian contribution is supported by the Italian Space Agency (Agenzia Spaziale Italiana, ASI) through contract ASI-OHBI-2022-13-I.0, agreements ASI-INAF-2022-19-HH.0 and ASI-INFN-2017.13-H0, and its Space Science Data Center (SSDC) with agreements ASI-INAF-2022-14-HH.0 and ASI-INFN 2021-43-HH.0, and by the Istituto Nazionale di Astrofisica (INAF) and the Istituto Nazionale di Fisica Nucleare (INFN) in Italy.  This research used data products provided by the IXPE Team (MSFC, SSDC, INAF, and INFN) and distributed with additional software tools by the High-Energy Astrophysics Science Archive Research Center (HEASARC), at NASA Goddard Space Flight Center (GSFC).
This work made use of data supplied by the UK Swift Science Data Centre at the University of Leicester.

\end{acknowledgments}

%
\facilities{IXPE, Swift}

\software{
\textsc{HEAsoft} \citep{HEAsoft},
\textsc{ixpeobssim} \citep{Baldini2022},
\textsc{Xspec} \citep{xspec},
\textsc{ds9} \citep{ds9},
\textsc{AstroPy} \citep{astropy2013,astropy2018,astropy2022}
}

\bibliography{tevbinary}
\bibliographystyle{aasjournal}

\end{document}